%% file: main.tex
\documentclass{bmvc2k}

\usepackage{bm}
\usepackage{float} 
\usepackage{xcolor}
\usepackage{comment}
\usepackage{overpic}
\usepackage{amsmath}
\usepackage{amssymb}
\usepackage{bbding}
\usepackage{array}
\usepackage{mathrsfs}
\usepackage{amsfonts}
\usepackage{graphicx}
\usepackage{appendix}
\usepackage{footnote}
\usepackage{booktabs}
\usepackage{wrapfig}
\usepackage{multirow}
\usepackage{verbatim}
\usepackage[T1]{fontenc}
\usepackage[mathscr]{eucal}

\newcommand{\model}[0]{{\textcolor{black}{K-Space Transformer}}}

\title{K-Space Transformer for Undersampled MRI Reconstruction}

\addauthor{Ziheng Zhao}{Zhao_Ziheng@sjtu.edu.cn}{1}
\addauthor{Tianjiao Zhang}{xiaoeyuztj@sjtu.edu.cn}{1,2}
\addauthor{Weidi Xie$^\dagger$}{weidi@sjtu.edu.cn}{1,2}
\addauthor{Yanfeng Wang$^\dagger$}{wangyanfeng@sjtu.edu.cn}{1,2}
\addauthor{Ya Zhang}{ya_zhang@sjtu.edu.cn}{1,2}

\addinstitution{
 Cooperative Medianet Innovation Center\\
 Shanghai Jiao Tong University\\
 Shanghai, China
}
\addinstitution{
 Shanghai AI Laboratory\\
 Shanghai, China
}

\runninghead{Zhao ET AL}{K-Space Transformer for Undersampled MRI Reconstruction}

\begin{document}

\maketitle

\def\thefootnote{}\footnotetext{$\dagger$ corresponding authors.}
\def\thefootnote{\arabic{footnote}}

\begin{abstract}
This paper considers the problem of undersampled MRI reconstruction. We propose a novel Transformer-based framework for directly processing signal in k-space, 
going beyond the limitation of regular grids as ConvNets do.
We adopt an implicit representation of k-space spectrogram, treating spatial coordinates as inputs, 
and dynamically query the sparsely sampled points to reconstruct the spectrogram, {\em i.e.}~learning the inductive bias in k-space.
To strike a balance between computational cost and reconstruction quality, 
we build the decoder with hierarchical structure to generate low-resolution and high-resolution outputs respectively. To validate the effectiveness of our proposed method, we have conducted extensive experiments on two public datasets, 
and demonstrate superior or comparable performance to state-of-the-art approaches. 
Project page: \href{https://zhaoziheng.github.io/Website/K-Space-Transformer}{https://zhaoziheng.github.io/Website/K-Space-Transformer}

\end{abstract}

%-------------------------------------------------------------------------
\input{intro-bmvc}
\input{RelatedWork}

\input{method-bmvc}
\input{experiment-bmvc}
\input{conclusion-bmvc}
\input{Ack}

\clearpage
\input{Appendix}

\end{document}

%% file: intro-bmvc.tex
\section{Introduction}
\label{sec:intro}

Magnetic Resonance Imaging (MRI) has been widely adopted as an efficient and non-invasive approach for routine examination and diagnosis.
In general, the signal is collected, digitized and plugged into k-space, {\em i.e.} an array of numbers representing spatial frequencies in the MR image, and then inverse fourier transformed to derive the MR image. 
However, due to hardware constraint, the full signal acquisition in k-space is time-consuming, 
which is uncomfortable and cause artifacts due to patient or physiological motions.
To alleviate this problem, various techniques have been developed for MR image reconstruction from undersampled k-space signal.

% For the last two decades,  
% compressed sensing~(CS) has been the major breakthrough in this field,
% allowing to reconstruct k-space signals from only partial measurements,
% however, the iterative optimization process often suffers from heavy parameter tuning~\cite{2017Deep} and long processing time~\cite{2015Reducing}.
In recent years, deep learning has been widely adopted for its better reconstruction performance and real-time imaging over traditional CS-based methods~\cite{zeng2021review}.
Generally speaking, existing works can be cast into two paradigms,
one focuses on reconstruction in image domain~\cite{UNet4MRI,D5C5,DAGAN,OUCR},
with the k-space information only being used in data consistency layer or loss function.
The other line of research takes k-space reconstruction into consideration, 
for example, integrate it to the normal image reconstruction pipeline in a recurrent or parallel manner~\cite{KIKINet,HybridCascade,DuDoRNet,CDFNet}; 
Or perform reconstruction purely in k-space~\cite{ksl2019,acnn2021}.
% and solve it with  unique characteristics such as low-rankness~\cite{loraks2013}.

In terms of the computational architecture,
almost all the existing approaches base k-space reconstruction on convolutional neural networks (CNNs), 
with two inevitable limitations: 
{\em First}, 
the inductive bias of CNNs is fundamentally not suitable for processing k-space spectrogram,
for example, the kernels in CNNs are normally shared across all spatial positions, 
{\em i.e.}~equivariance property, however, on k-space spectrogram, 
spatial positions stand for the frequency bins of the sine and cosine functions,
same patterns occurring at different positions may refer to completely different information;
{\em Second}, CNNs exploit spatial locality by limiting the connectivity between neurons to adjacent regions, ending up small receptive field. 
Although this problem can be alleviated by pooling or cascading more layers, 
additional computations are introduced.
It thus remains unclear what neural network architecture can capture the suitable inductive bias for signal in k-space.

\begin{figure}[t] 
\centering 
\includegraphics[width=\textwidth]{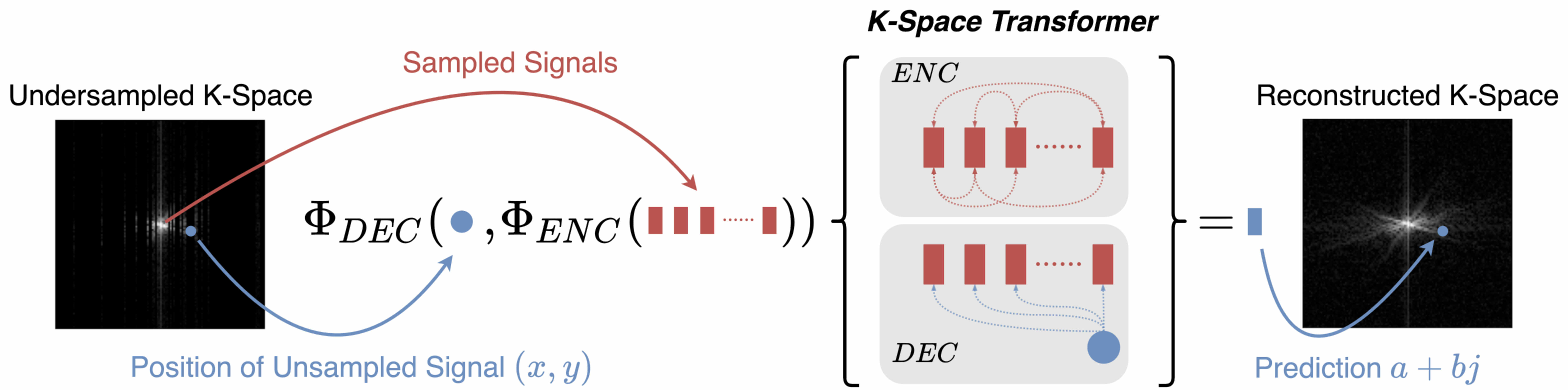}
\vspace{-10pt}
\caption{We adopt a transformer for implicit representation of k-space spectrogram (termed as K-Space Transformer).}
\vspace{-10pt}
\label{fig:teaser} 
\end{figure}

%while basic CNNs with weight-sharing kernel are inherently insensitive to such position information\cite{CoordConv}. 
%which contradicts the globality prior in frequency domain.

In this paper, 
we propose a novel Transformer-based architecture for MRI reconstruction,
termed as \model, that enables to learn inductive bias in k-space, 
as shown in Figure~\ref{fig:teaser}.
Inspired by the implicit representation~\cite{LIIF,NERF}, 
we treat the coordinates of unsampled k-space points as inputs to the Transformer Decoder,
iteratively query information from those sampled points, 
and model relationships between any frequency bins beyond the regular spatial grids.
Thanks to the high flexibility of such representation, 
we design a hierarchical decoder with coarse-to-fine reconstruction strategy, to strike a balance between performance and computational cost. 
Since pure k-space reconstruction would ignore important spatial bias in MRI, we further introduce an image domain refinement module to solve this problem.
To summarise, our main contributions are three-fold: 
\begin{itemize}
    \item We introduce a novel hierarchical transformer architecture that aims to learn suitable inductive bias for directly processing signal in k-space. To the best of our knowledge, we are the first to introduce transformer to k-space reconstruction.
    \item We integrate image domain refinement module into the transformer decoder, and prove it to be complementary to the pure k-space reconstruction.
    \item We validate \model~on two public datasets, demonstrating comparable or superior performance than previous state-of-the-art approaches.
\end{itemize}

%% file: RelatedWork.tex
\section{Related Work}
\paragraph{Undersampled MRI Reconstruction} is an ill-posed inverse problem. Based on the low dimensionality prior, traditional CS-Based methods generally leverage sparsity regularization in certain domain to reconstruct the image in iterative ways~\cite{MRI-CS}. 
However, the iterative optimization process often suffers from heavy parameter tuning~\cite{2017Deep} and long processing time~\cite{2015Reducing}. Some works extend these approaches by construct deep networks to replace handcrafted parameters or functions~\cite{DeepADMM,ISTA}. Another popular school of thought uses deep networks to learn a direct mapping from input to output~\cite{ksl2019,acnn2021,KIKINet,HybridCascade,DuDoRNet,CDFNet,UNet4MRI,D5C5,DAGAN,OUCR}. Most of these works focus on image domain reconstruction. For example, Schlemper et al.~\cite{D5C5} view the problem as image de-aliasing and propose a deep cascade framework of CNNs. Works in this direction have exploited various network architectures, including U-Net~\cite{UNet4MRI}, GAN~\cite{DAGAN,GANCS}, convolutional RNN~\cite{OUCR}, Transformer~\cite{SwinMR} etc. Zhu et al.~\cite{zhu2018image} propose to learn a mapping from k-space to image domain through fully connected layers, followed by convolutional layers. By contrast, Han et al.~\cite{ksl2019} use U-Net for direct k-space interpolation, complete the reconstruction purely in k-space. Du et al.~\cite{acnn2021} further integrate channel and spatial attention mechanism into k-space convolution. Dual domain reconstruction is another natural choice to combine knowledge from both image domain and k-space~\cite{KIKINet,HybridCascade,DuDoRNet,CDFNet}. While k-space knowledge is commonly considered, the works so far base k-space learning on CNNs. \\[-0.7cm]

\paragraph{Implicit Neural Representation} parameterizes signals as continuous functions with neural networks, mapping coordinates to values. It have been widely applied to 3D object and scene modelling~\cite{IM-NET,SAL, eslami2018neural,jiang2020local,sitzmann2019scene,NERF}, and image processing~\cite{StyleGAN3,shaham2021spatially,LIIF}. However, its potential in medical image reconstruction is relatively less exploited. Wu et al.~\cite{wu2021irem} learn continuous volumeric function from discrete MRI slices to reconstruct high resolution 3D images. 
Shen et al.~\cite{shen2022nerp} takes a similar strategy but further embed prior image information into the network. 
Another related study~\cite{sun2021coil} uses MLP to represent the measurement field of a CT image, and reconstruct from the generated measurements. 
Our work differs from the above mentioned works, as we share the function space across different instances, instead of optimizing unique representation for each individual object, and thus learn general prior knowledge of k-space. 

% However, its potential in medical domain is relatively less exploited. Zhang et al.~\cite{NeRD} propose to optimize underlying functions mapping image coordinates to feature distribution in segmentation task. Juhl et al.~\cite{juhl2021implicit,sorensen2022nudf} propose to learn neural distance field of 3D anatomy, which facilitates anatomy classification and segmentation. As for medical image reconstruction, 

%% file: method-bmvc.tex
\section{Method}

In this paper, 
we consider the problem of undersampled MRI reconstruction,
with the goal of learning a function that maps the under-sampled spectrogram to high-quality MR images, 
$\mathcal{I} = \mathrm{\Phi}(x_s; \Theta)$, 
where $\mathcal{I}, x_s$ refers to the output MR image,  and the input under-sampled spectrogram respectively,
$\Theta$ denotes the set of learnable parameters.

Unlike existing work on MRI reconstruction
that generally exploit ConvNets in {\em image} or {\em k-space}, 
we adopt the implicit representation, and propose a novel Transformer-based architecture for learning the inductive bias in {\em k-space}.
In the following sections,
we start by introducing the fundamental building blocks, 
namely, the Encoder module~($\Phi_{\text{ENC}}$) that learns a compact feature representation from sampled k-space points;
and Decoder module~($\Phi_{\text{DEC}}$) that reconstructs MRI by alternating k-space decoding and image domain refinement.
After that, we describe in detail how these blocks are used for constructing an hierarchical model,
that strikes a balance on the performance and computational complexity trade-off.

%As shown in Figure~\ref{fig:arch},
%our proposed model consists of three components, 
%the Encoder~($\Phi_{\text{ENC}}$) learns a compact feature representation from the sampled frequency bins;
%the Decoder~($\Phi_{\text{DEC}}$) that alternates K-space reconstruction and refinement in image domain. }

%coarsely based on the feature representation; 
%and a High-Resolution Decoder $\Phi_{\text{HRD}}$ further refines the reconstruction, 
%outputting higher resolution. }

%At an high-level, as shown in Figure~\ref{fig:arch}, our proposed architecture $\Phi(\cdot)$ can be formulated as:
% \begin{align*}
%    & \Phi(\cdot) = \Phi_{\text{HRD}} \circ \Phi_{\text{LRD}} \circ \Phi_{\text{ENC}} (\cdot)
%\end{align*}

%efforts that relay on CNN and thus retain the 2D structure of the spectrogram, we formulate this problem from a novel view: the under-sampled k-space spectrogram is represented by a sequence of the sampled frequencies, from which we need to decode the unknown ones. 

\begin{figure}[t] 
\centering 
\includegraphics[width=\textwidth]{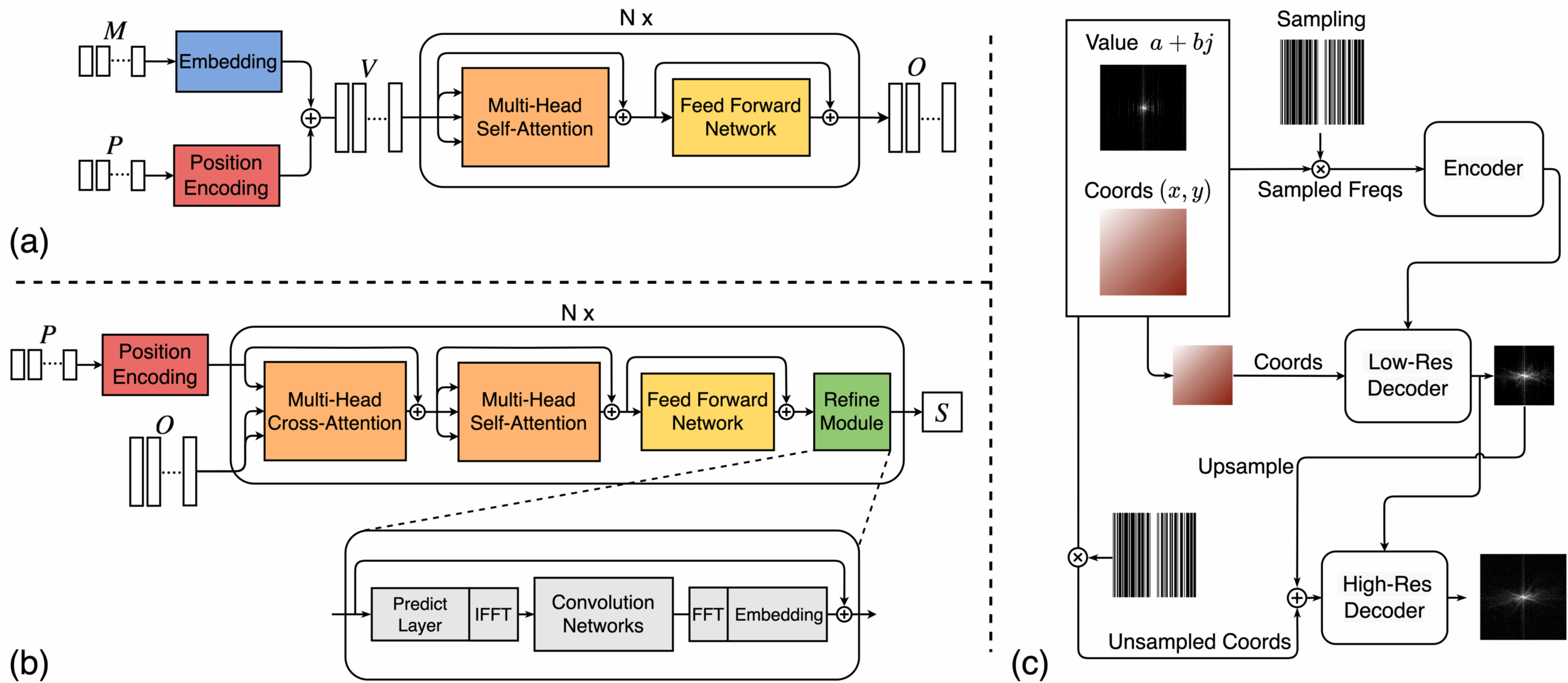}
% \vspace{-10pt}
\caption{(a) The structure of Encoder; (b) The structure of Decoder; (c) The overall framework of the K-Space Transformer.}
% \vspace{-10pt}
\label{fig:arch} 
\end{figure}

\subsection{Feature Encoder}
\label{sec:enc}

Here, the goal is to compute a compact feature representation for the sampled points in k-space. 
In detail, given a set of $n$ sampled points on spectrogram, 
{\em i.e.}~$\{s_i, \dots, s_n\}$, 
with $s_i =(m_i, p_i)$ refers to a combination of sampled complex value~($m_i \in \mathbb{R}^2$) 
and 2D spatial positions~($p_i \in \mathbb{R}^2$).
Note that, $n$ is usually a small subset of all points on a spectrogram~($\mathbb{R}^{W \times H}$).
As a \textbf{tokenisation} procedure that converts the sampled points into a vector sequence,
$\mathcal{V} = \{v_1, v_2, \dots, v_n\} \in \mathbb{R}^{n \times d}$, with
\begin{align}
\label{eq:tok}
    v_i = \mathrm{\Phi}_{\textsc{Tokenize}}(s_i) = \text{MLPs}(m_i) + \text{PE}(p_i)
\end{align}
where one Multilayer Perceptrons~(MLPs) layer is applied to the sampled value,
and \textbf{PE} refers to positional encodings with sine and cosine functions.

As shown in Figure~\ref{fig:arch}~(a), 
the Encoder includes multiple standard transformer encoder layers,
consisting of a multi-head self-attention module~(MHSA), 
a feed forward network (FFN) and residual connections.
Through self-attention module, 
the global dependency between each sampled point are captured, 
and the FFN further enriches the feature representation.
\begin{equation}
    O = \mathrm{\Phi}_{\textsc{K-Enc}}(\mathcal{V}) \in \mathbb{R}^{n \times d}
\end{equation}
where $O$ denotes the output from encoder, 
with same dimensions as input sequence vectors.
For more details, 
we would refer the readers to the original Transformer paper~\cite{Vaswani17}.

\subsection{Feature Decoder}
\label{sec:decoder}

In this section, 
we describe the decoding procedure that maps the encoded representation to MR images. 
We adopt an hybrid architecture that alternates between {\em k-space} decoding, 
and refinement in {\em image domain}, shown in Figure~\ref{fig:arch}~(b). \\[-0.7cm]

\paragraph{Decoding in k-space. }
We adopt standard Transformer decoder layers, 
where the \texttt{Key} and \texttt{Value} are generated by applying two different linear transformations on encoder's outputs,
and the normalised positional coordinates for desired output~($\mathbb{R}^{W \times H}$) are encoded as \texttt{Query}:
\begin{align}
\label{equ:query}
& q_i = W^q \cdot \text{PE}(p_i) \text{\hspace{10pt}} \forall p_i \in \mathbb{R}^{WH \times 2} \\
& k_i, \text{\hspace{5pt}} v_i = W^k \cdot O_i, \text{\hspace{5pt}} W^v \cdot O_i \text{\hspace{10pt}} \forall i \in [1,n] \\
& S = g(\Phi_{\textsc{K-Dec}}(Q, K, V)) \in \mathbb{R}^{W \times H \times 2}
\end{align}
where \textbf{PE} refers to positional encodings with sine and cosine functions,
$q_i, k_i, v_i$ refer to $i$-th vector of $Q, K, V$ respectively, 
and $S$ denotes the predicted spectrogram by
applying a linear MLP layer~($g(\cdot)$) on the output from transformer decoders~($\Phi_{\textsc{K-Dec}}(\cdot)$), consisting of multi-head self-attention~(MHSA), 
multi-head cross-attention~(MHCA), a feed-forward network~(FFN), and residual connections.
Note that, the resolution of the reconstructed spectrogram can be controlled by granularity of {\em query}, {\em i.e.}~the number of positional coordinates.

Thanks to the flexibility of Transformer decoder layers, 
dependencies between all points on spectrogram can be effectively captured, thus enable to learn the inductive bias in k-space.
However, one issue is, 
the point-wise update in k-space will globally affects all pixels in image domain, 
disregarding the spatial bias in MRI, that is, {\em structures tend to be roughly aligned}. 
To resolve such issue, we alternate the processing in {\em k-space} and {\em image domain}.\\[-0.7cm]

\paragraph{Refinement in Image Domain. }
Here, we convert the reconstructed spectrogram into {\em image domain} with differentiable inverse Fast Fourier Transformer~(FFT), and process the images with standard convolution layers~($\Phi_{\textsc{Img-Refine}}(\cdot)$):
\begin{align}
    \hat{S} = \mathcal{T}(\Phi_{\textsc{Img-Refine}}(\mathcal{T}^{-1}(S)))
\end{align}
where $\mathcal{T}$ and $\mathcal{T}^{-1}$ refer to FFT and inverse FFT respectively,
the output is $\hat{S} \in \mathbb{R}^{W\times H \times 2}$. \\[-0.7cm]

% Note we fuse the refinement results with the dense feature of k-space decoding, which is different from existing dual-domain approaches\cite{KIKINet,HybridCascade,DuDoRNet} that discard feature maps from previous blocks when alternating the reconstruction in dual domains. We finds preserving the dense feature through decoder is critical to increase the reconstruction quality in deeper layers. \\[-0.7cm]

\paragraph{Discussion: } 
At an high level, instead of treating the spectrogram as a look-up table array, 
{\em i.e.}~values on regular grids, 
\textbf{we adopt the implicit representation, learning a continuous function that maps the input positional coordinates to values, conditioned on the visible k-space samples}.
Such representation is highly flexible,
as it not only enables to explicitly model the dependencies between different frequency bins,
but also support to output spectrogram of different resolutions by simply varying the density of query coordinates.

%resembling a similar idea to the implicit neural representation~(\weidi{give citation}),

%\weidi{reword the following sentence, to medical imaging audience, it may not be clear what does it mean.}
%it can be viewed as a conditional continuous function representing the entire spectrogram. 
%It thus has no restriction from a fixed resolution, offering a lot flexibility for practical implementation.

\subsection{Hierarchical Decoder}
\label{sec:hierarchy}
With the basic building blocks introduced, 
we construct an hierarchical decoder, to strike a balance between the computational cost and performance trade-off.
In particular, 
we adopt a low-resolution~(LR) decoder and a high-resolution~(HR) decoder to guide the reconstruction with progressive resolutions, and tailor them for different purposes:
$\mathrm{\Phi}_{\text{DEC}}(\cdot) = \mathrm{\Phi}_{\text{HRD}} \circ \mathrm{\Phi}_{\text{LRD}}(\cdot)$,
where $\mathrm{\Phi}_{\text{LRD}}$ aims to reconstruct the spectrogram on a lower resolution, 
focusing on the overall anatomical structure. 
As shown in Figure~\ref{fig:arch}~(c), this is simply implemented by adapting the normalised positional coordinate $p_i$ in Equation~\ref{equ:query} to a down-sampled spectrogram. 
Here, we {\bf do not} use image domain refinement,
and only implement the LR decoder with standard Transformer decoder layers. 

The HR decoder aims to fill the texture details based on the LR output. 
We accordingly map the tokenised HR coordinates~(Eq.~\ref{equ:query}) to queries, 
and treat the two projections of LR decoder output as \texttt{Key} and \texttt{Value}. 
To alleviate the computation from self-attention module, 
we simplify each decoder layer in HR decoder with {\bf only cross-attention module left}.\\[-0.7cm]

\paragraph{Discussion: }
As for computational complexity, 
when decoding the $m$-length sequence from the $n$-length sampled points
with a standard Transformer decoder, 
the complexity would be $\mathcal{O}(mn)$ for MHCA and $\mathcal{O}(m^2)$ for MHSA, 
which is prohibitively expensive when $m$ is very large. 
However, 
in our case, 
by constructing the hierarchical model, 
the computation complexity is reduced to $\mathcal{O}(ml+nl+l^2)$,
as the MHSA can simply be done in LR decoder, 
drastically cutting the memory consumption, 
when $l$ is set to be much smaller than $m$.

%% file: experiment-bmvc.tex
\section{Experiment}

\subsection{Dataset \& Metrics}

We employ two datasets for evaluation, namely, OASIS and fastMRI.
OASIS~\cite{OAS} contains 3,398 single-coil T1-weighted brain MRI volumes, 
each has 175 slices. 
We split them into train and test sets with ratio 3:1 and uniformly draw 10\% from them. Additionally, fastMRI has 1172 single-coil PD-weighted knee MRI volumes, each includes 35 slices. 
We use 973 for training, 199 for testing and uniformly draw 25\% slices from both. Following conventional evaluation, 
we experiment on five undersampling settings: 
Gaussian 2D sampling with 10$\times$, 5$\times$ and $2.5\times$ accelerations; Uniform 1D sampling with 5$\times$ and $2.5\times$ accelerations. 
Peak signal to noise ratio~(PSNR) and structural index similarity~(SSIM) are used as metrics.

% \subsection{Implementation Details}
% Our K-Space Transformer is implemented with 4 Encoder layers, 4 Low-Resolution Decoder layers and 6 High-Resolution Decoder layers. Following previous works~\cite{acnn2021,OUCR,ksl2019,D5C5,HybridCascade}, we divide the complex MR signals into real and imaginary channels and enforce $\ell_2$ loss in image domain. Besides, to accelerate training, we also apply deep supervision on each layer of decoders. We adopt AdamW optimizer and cosline annealing schedule with initial learning rate of $5\times10^{-4}$. Training is conducted on 4 RTX 3090. For more details, please refer to our supplementary materials.
% \input{./tab/tab1.tex}
%\zzh{tab3.tex divide results into two tables by mask}

\subsection{Comparison to State-of-the-art}
\label{sec:baseline}
We compare to 6 representative methods, 
including UNet~\cite{UNet} for image domain reconstruction and k-space reconstruction, 
Deep ADMM Net~\cite{DeepADMM}, D5C5~\cite{D5C5}, SwinMR\cite{SwinMR}
and previous state-of-the-art OUCR~\cite{OUCR}. For implementation details of K-Space Transformer and the baseline methods, please refer to the supplementary details. \\[-0.6cm]

\paragraph{Quantitative Results: }
As shown in Table~\ref{tab:uniform}, 
our method achieves the best results under all 1D settings. 
On OASIS, we improved the PSNR from 29.52 dB to 31.50 dB with 5$\times$ acceleration, 
and for 2.5$\times$ sampling, our improvement is even enlarged to 2.57 dB. 
On fastMRI, we outperform the best baseline slightly,
however we also notice that the top approaches perform quite closely, 
we conjecture this is because the dataset is quite noisy, 
which may lead to an unprecise supervision for training and evaluation. 
In addition, 
it's noticeable that K-UNet leads to poorer performance compared with UNet, 
showing that na\"ive CNN-based methods {\bf do not} give suitable inductive bias for k-space reconstruction.
On 2D settings~(Table~\ref{tab:gaussian}), 
our method either outperforms or achieve competitive results to the state-of-the-part approaches. \\[-0.3cm]

\input{baseline-tab.tex}

\begin{figure}[!htb]
    \centering
    \includegraphics[width=\textwidth]{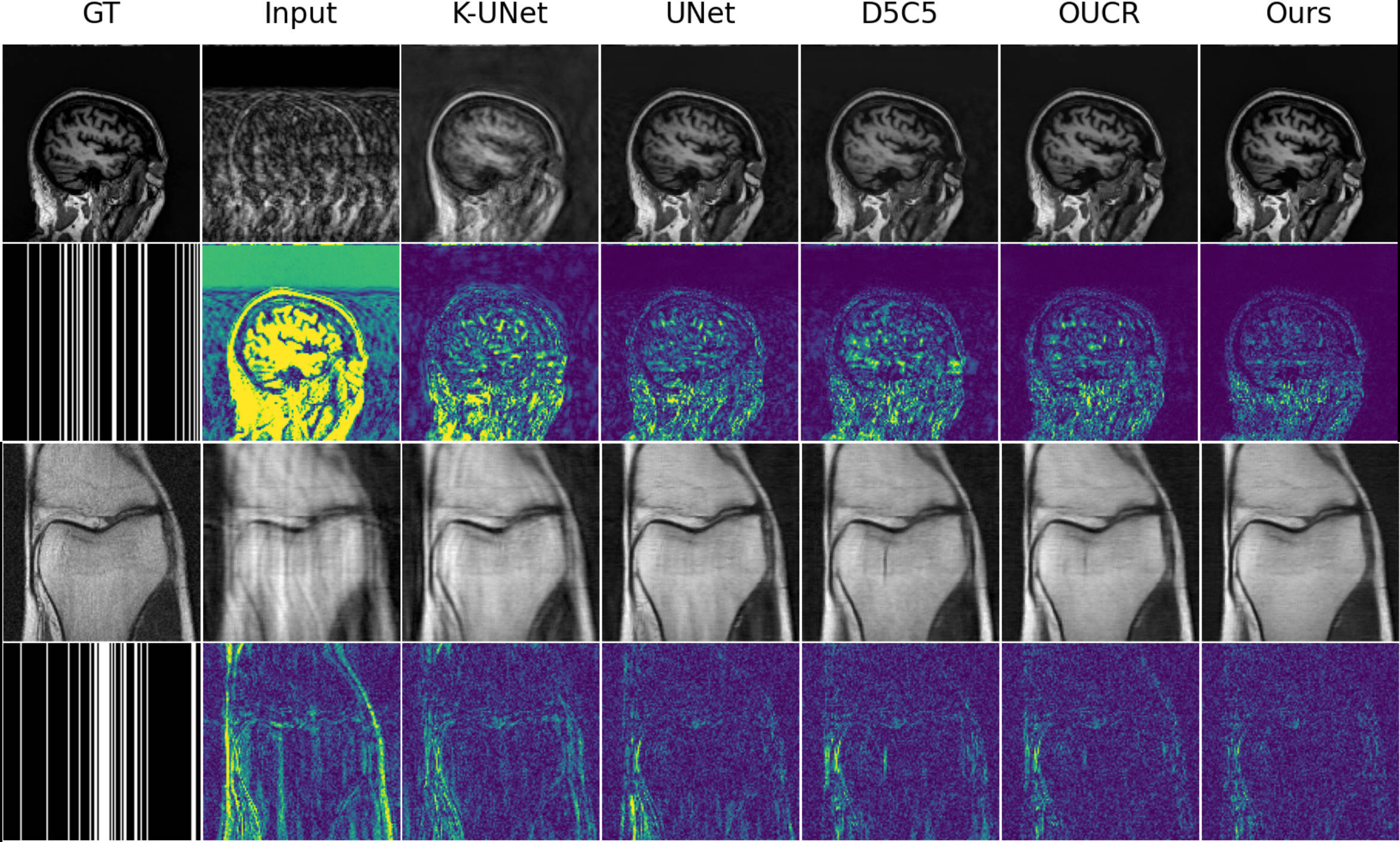}
    \vspace{-15pt}
    \caption{Qualitative comparison under 5$\times$ uniform sampling setting. The even-numbered rows show the sampling trajectories and error maps, where brighter means higher error.}
    \label{fig:visual}
    %\vspace{-10pt}
\end{figure}

\paragraph{Qualitative Results: }
In Figure~\ref{fig:visual},
we show some predictions from \model ~on 5$\times$ uniform sampling setting. 
On OASIS, our method clearly retain more texture information and less structural loss compared with others, suggesting that our method is more robust for such severe distortion in image domain.
While on fastMRI, the groundtruth images are often noisy,
which might explain why the top approaches perform similarly. Due to space limit, we attach more qualitative comparison results in the supplementary material.
%Due to space limit, we attach more inference results of other experiments in appendix. 

\vspace{-.3cm}
\subsection{Ablation Study}
\label{sec:abl}

To investigate the effectiveness of hybrid learning and the hierarchical structure, we remove the image domain refinement module~(RM), and the LR decoder~(LRD) sequentially on OASIS dataset.

As shown in Table~\ref{tab:2}, 
without the refinement module, the reconstruction quality drops significantly,
showing the image domain refinement can indeed provide supplementary information for k-space reconstruction. Nevertheless, our method still achieves competitive
\begin{wraptable}{r}{8cm}
% \vspace{-.3cm}
\setlength\tabcolsep{8pt}  
\centering
\resizebox{.58\textwidth}{!}{%
\begin{tabular}{cc|cc|cc}
\hline
\multicolumn{2}{c|}{Modules} & \multicolumn{2}{c|}{Gaussian 2D 5$\times$} &  \multicolumn{2}{c}{Uniform 1D 5$\times$} \\ \hline
\ LRD \ & \ RM \ & PSNR & SSIM & PSNR & SSIM \\ \hline
\Checkmark & \Checkmark & 37.47 & 0.9823 & 31.50 & 0.9528 \\ 
\Checkmark & \XSolidBrush & 32.97 & 0.9337 & 27.52 & 0.8876 \\ 
\XSolidBrush & \XSolidBrush & 28.53 & 0.8416  & 21.31 & 0.6114 \\ \hline
\end{tabular}
}
\vspace{5pt}
\caption{Quantitive ablation study on critical designs.}
\label{tab:2}
\vspace{-5pt}
\end{wraptable}
 results and outperforms K-UNet by a large margin (refer to Table~\ref{tab:uniform} and~\ref{tab:gaussian}),
which justifies Transformer's superiority for k-space reconstruction. 
In addition, removing the LR decoder further leads to an obvious performance degradation, 
implying that our hierarchical design also plays an essential role
in guaranteeing an excellent reconstruction quality with a moderate computational cost.

\begin{figure}[htbp]
    \centering
    \includegraphics[width=\textwidth]{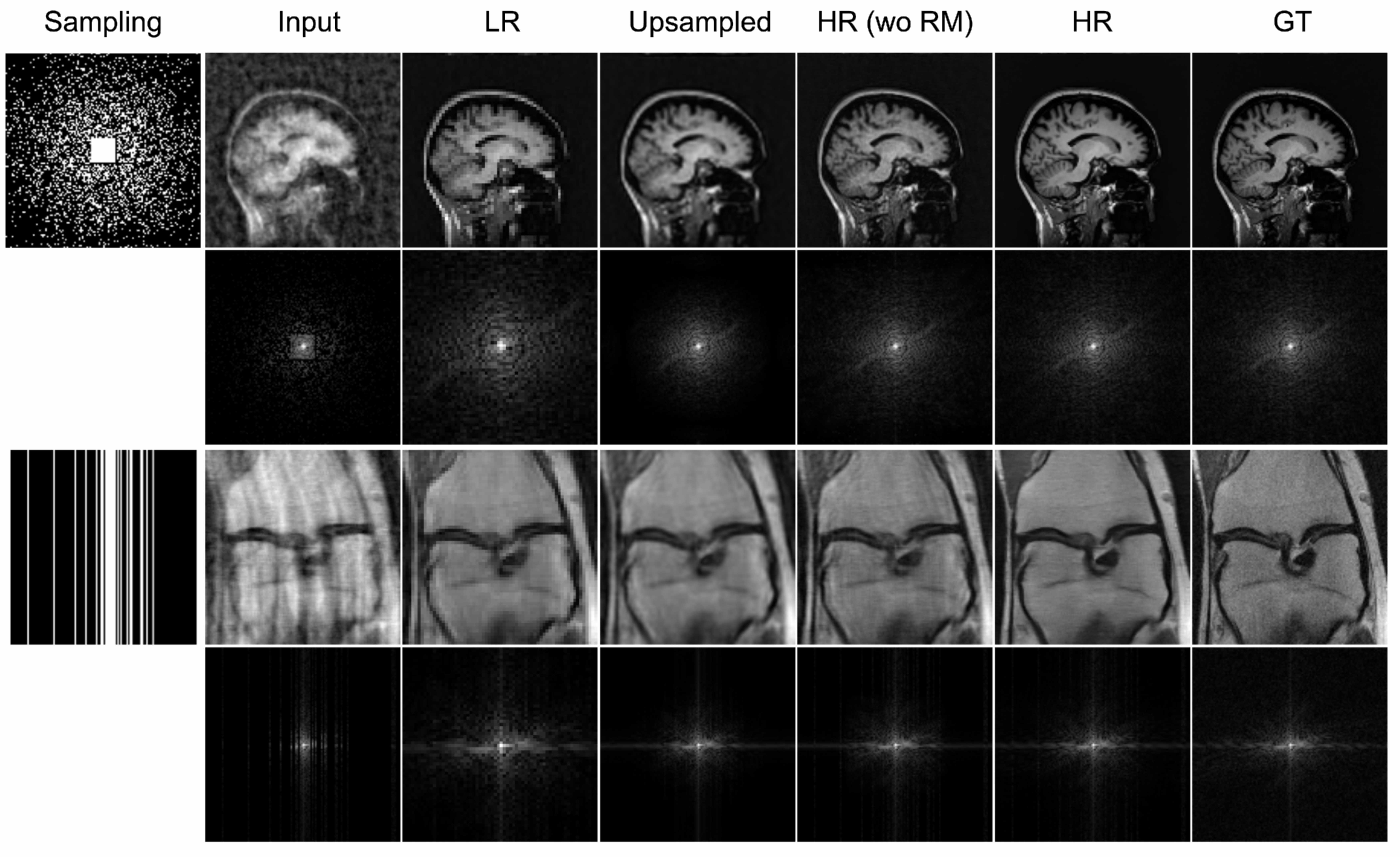}
    \vspace{-15pt}
    \caption{Intermediate results of K-Space Transformer: LR denotes the reconstruction output of LR decoder; Upsampled denotes the up-sampled result; HR and HR (wo RM) refer to the output of HR decoder and that without refinement module.}
    \label{fig:inter}
    \vspace{-10pt}
\end{figure}

To give a more distinct illustration, 
we further visualize some intermediate results of \model~in Figure~\ref{fig:inter}. 
We observe the LR decoder capable to reconstruct a basic outline of the structure, 
equally the low-mid frequency bands of the spectrogram. 
Based on the upsampled rough result, the HR decoder fills in the texture details, 
{\em i.e.}, completing higher frequency bands. 
The coarse-to-fine progress is consistent with our assumption in Section~\ref{sec:hierarchy}. 
Compared to standard HR decoder, those without refinement module contain more noise over the structures and background, which is easier to remove via spatial convolution. 
It indicates the inductive bias of image domain are complementary to k-space. \\ [-0.4cm]

\paragraph{On Positional Encoding and Receptive Field.}
To make comparison between UNet and Transformer, 
we also extend the standard U-Net with additional positional encoding and large receptive field~(more parameters). By explicitly concatenate the postional encodings with k-space spectrogram as the input, we change the convolution into position dependent~(denoted as U-Net P); 
We further enlarge the receptive filed by adopting $5\times5$ kernel and deeper convolution layers~(denoted as K U-Net P L). 

As shown in Table~\ref{tab:UNetAblation},  
both positional encoding and large receptive field improve the performance of a standard U-Net. 
However, the gain is marginal, note that a large receptive field would increase the model size exponentially.

\begin{table}[!htb]
\setlength\tabcolsep{8pt}  
\centering
\resizebox{.9\textwidth}{!}{
\begin{tabular}{c|cccc|cccc}
\hline
\multirow{3}{*}{Method} & \multicolumn{4}{c|}{OASIS} & \multicolumn{4}{c}{fastMRI} \\ \cline{2-9} & \multicolumn{2}{c|}{Uniform} & \multicolumn{2}{c|}{Gaussian} & \multicolumn{2}{c|}{Uniform} & \multicolumn{2}{c}{Gaussian} \\ 
\cline{2-9} & \multicolumn{1}{c|}{5$\times$} & \multicolumn{1}{c|}{2.5$\times$} & \multicolumn{1}{c|}{5$\times$} & 2.5$\times$ & \multicolumn{1}{c|}{5$\times$} & \multicolumn{1}{c|}{2.5$\times$} & \multicolumn{1}{c|}{5$\times$} & 2.5$\times$ \\
\hline
K U-Net & \multicolumn{1}{c|}{24.53} & \multicolumn{1}{c|}{26.50}  & \multicolumn{1}{c|}{28.76} & 33.96 & \multicolumn{1}{c|}{24.68} & \multicolumn{1}{c|}{26.99} & \multicolumn{1}{c|}{27.15} & 28.90 \\ 
\hline
K U-Net P & \multicolumn{1}{c|}{24.55} & \multicolumn{1}{c|}{26.67} & \multicolumn{1}{c|}{29.18} & 35.10 & \multicolumn{1}{c|}{24.70} & \multicolumn{1}{c|}{28.13} & \multicolumn{1}{c|}{27.56} & 29.98 \\
\hline
K U-Net P L & \multicolumn{1}{c|}{24.64} & \multicolumn{1}{c|}{27.00} & \multicolumn{1}{c|}{29.87} & 35.18& \multicolumn{1}{c|}{24.71} & \multicolumn{1}{c|}{28.44} & \multicolumn{1}{c|}{27.68} & 30.01 \\ 
\hline
Ours & \multicolumn{1}{c|}{31.50} & \multicolumn{1}{c|}{36.32} & \multicolumn{1}{c|}{37.47} & 43.68 & \multicolumn{1}{c|}{27.80} & \multicolumn{1}{c|}{31.04} & \multicolumn{1}{c|}{29.88} & 31.97 \\ 
\hline
\end{tabular}
}
\vspace{5pt}
\caption{Comparision between U-Net variants and Transformer (PSNR).}
%\vspace{-1cm}
\label{tab:UNetAblation}
\vspace{-13pt}
\end{table}

%The whole coarse-to-fine reconstruction progress is exactly consistent with our assumption in section \ref{sec:hierarchy}. 

\subsection{Visualising Transformer in K-Space Learning}

\begin{figure}[!htb]
\centering
\includegraphics[width=0.98\textwidth]{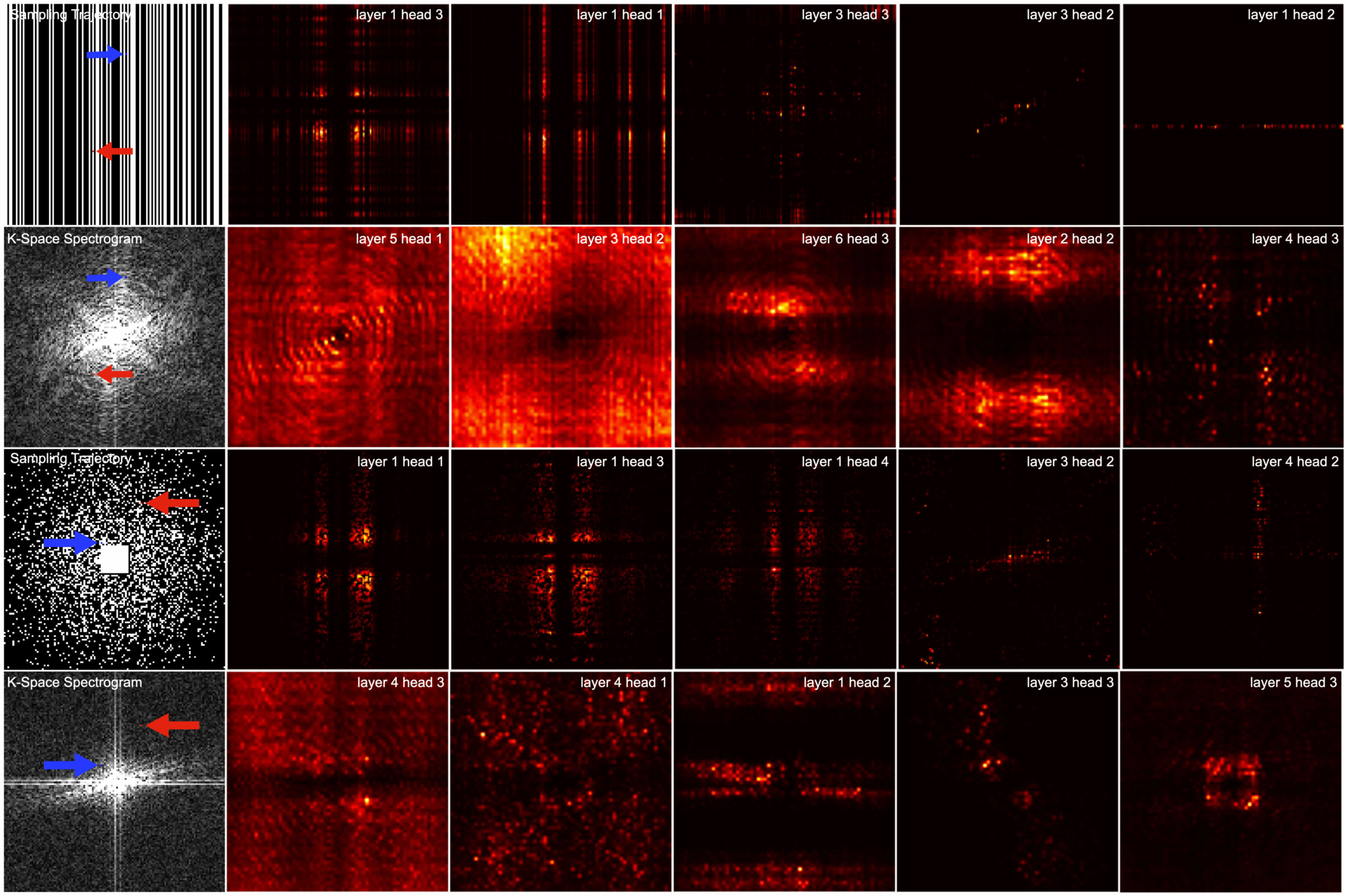}
    \vspace{-5pt}
    \caption{Attention maps from the encoder and decoder under different sampling patterns. We randomly select a sampled point (indicated by the red arrow) and visualize its attention map in encoder in the upper rows (i.e. relationship with other sampled points); We randomly select a unsampled point(the blue arrow) and visualize its attention map in HR decoder in the lower rows (i.e. relationship with the reconstructed LR spectrogram).} 
    \label{fig:attention}
    \vspace{-10pt}
\end{figure}

Here, we visualize the attention maps of our encoder and decoder according to,
{\em i.e.}, $\text{attn} = \text{softmax}(QK^T/\sqrt{d})$,
where $Q$ and $K$ refer to the \texttt{Query} and \texttt{Key} indicated in section \ref{sec:enc} and \ref{sec:decoder}. Recall that we expect the K-Space Transformer to fully exploit the relationship between any points for reconstruction. As shown in Figure~\ref{fig:attention}, we do observe both the global and local interactions have been well exploited through the attention mechanism, 
as there are some heads clearly scanning the spectrogram globally, 
and others focusing on specific regions of interests,
implying that the model can indeed capture inductive bias in k-space. 

%% file: baseline-tab.tex
% \usepackage{multirow}
%\vspace{-15pt}
\begin{table}[!htb]
\centering
\setlength\tabcolsep{5pt}  
\resizebox{0.83\textwidth}{!}{
\begin{tabular}{c|cccc|cccc}
\hline
\multicolumn{1}{c|}{\multirow{3}{*}{Method}} & \multicolumn{4}{c|}{OASIS} & \multicolumn{4}{c}{fastMRI} \\ \cline{2-9} 

\multicolumn{1}{c|}{} & \multicolumn{2}{c|}{5 $\times$} & \multicolumn{2}{c|}{2.5 $\times$} & \multicolumn{2}{c|}{5 $\times$} & \multicolumn{2}{c}{2.5 $\times$} \\ \cline{2-9} 

\multicolumn{1}{c|}{} & \multicolumn{1}{c|}{PSNR} & \multicolumn{1}{c|}{SSIM} & \multicolumn{1}{c|}{PSNR} & \multicolumn{1}{c|}{SSIM} & \multicolumn{1}{c|}{PSNR} & \multicolumn{1}{c|}{SSIM} & \multicolumn{1}{c|}{PSNR} & \multicolumn{1}{c}{SSIM} \\ \hline

K-UNet &  \multicolumn{1}{c|}{24.53} & \multicolumn{1}{c|}{0.7808} & \multicolumn{1}{c|}{26.50} & 0.8483 & \multicolumn{1}{c|}{24.68} & \multicolumn{1}{c|}{0.6058} & \multicolumn{1}{c|}{26.99} & 0.7223 \\ \hline

Deep ADMM & \multicolumn{1}{c|}{26.03} & \multicolumn{1}{c|}{0.7202} & \multicolumn{1}{c|}{29.95} & \multicolumn{1}{c|}{0.8165} & \multicolumn{1}{c|}{24.43} & \multicolumn{1}{c|}{0.6172} & \multicolumn{1}{c|}{29.84} & \multicolumn{1}{c}{0.7838} \\ \hline

SwinMR & \multicolumn{1}{c|}{27.17} & \multicolumn{1}{c|}{0.9041} & \multicolumn{1}{c|}{29.83} & \multicolumn{1}{c|}{0.9435} & \multicolumn{1}{c|}{26.01} & \multicolumn{1}{c|}{0.6684} & \multicolumn{1}{c|}{29.10} & \multicolumn{1}{c}{0.7842} \\ \hline

UNet & \multicolumn{1}{c|}{27.53} & \multicolumn{1}{c|}{0.8895} & \multicolumn{1}{c|}{29.05} & 0.8988 & \multicolumn{1}{c|}{26.72} & \multicolumn{1}{c|}{0.6835} & \multicolumn{1}{c|}{29.81} & 0.7983 \\ \hline

D5C5 & \multicolumn{1}{c|}{27.66} & \multicolumn{1}{c|}{0.8895} & \multicolumn{1}{c|}{32.48} & 0.9615 & \multicolumn{1}{c|}{26.87} & \multicolumn{1}{c|}{0.6793} & \multicolumn{1}{c|}{30.64} & 0.8126 \\ \hline

OUCR & \multicolumn{1}{c|}{29.52} & \multicolumn{1}{c|}{0.9310} & \multicolumn{1}{c|}{33.75} & 0.9709 & \multicolumn{1}{c|}{27.71} & \multicolumn{1}{c|}{0.7009} & \multicolumn{1}{c|}{30.95} & 0.8168 \\ \hline

Ours & \multicolumn{1}{c|}{\textbf{31.50}} & \multicolumn{1}{c|}{\textbf{0.9528}} & \multicolumn{1}{c|}{\textbf{36.32}} & \textbf{0.9848} & \multicolumn{1}{c|}{\textbf{27.80}} & \multicolumn{1}{c|}{\textbf{0.7078}} & \multicolumn{1}{c|}{\textbf{31.04}} & \textbf{0.8189} \\ \hline
\end{tabular}}
\vspace{6pt}
\caption{Quantitative comparison on uniform 1D sampling settings.}
\label{tab:uniform}
\vspace{-12pt}
\end{table}

% \usepackage{multirow}
\begin{table}[!htb]
\centering
\setlength\tabcolsep{5pt}  
\resizebox{\textwidth}{!}{%
\begin{tabular}{c|cccccc|cccc}
\hline
\multirow{3}{*}{Method} & \multicolumn{6}{c|}{OASIS} & \multicolumn{4}{c}{fastMRI} \\ \cline{2-11} 
 & \multicolumn{2}{c|}{10 $\times$} & \multicolumn{2}{c|}{5 $\times$} & \multicolumn{2}{c|}{2.5 $\times$} & \multicolumn{2}{c|}{5 $\times$} & \multicolumn{2}{c}{2.5 $\times$} \\ \cline{2-11} 
 & \multicolumn{1}{c|}{PSNR} & \multicolumn{1}{c|}{SSIM} & \multicolumn{1}{c|}{PSNR} & \multicolumn{1}{c|}{SSIM} & \multicolumn{1}{c|}{PSNR} & SSIM & \multicolumn{1}{c|}{PSNR} & \multicolumn{1}{c|}{SSIM} & \multicolumn{1}{c|}{PSNR} & SSIM \\ \hline
%Input & \multicolumn{1}{c|}{20.99} & \multicolumn{1}{c|}{0.4856} & \multicolumn{1}{c|}{23.99} & \multicolumn{1}{c|}{0.5941} & \multicolumn{1}{c|}{26.92} & \multicolumn{1}{c|}{0.6936} & \multicolumn{1}{c|}{25.01} & \multicolumn{1}{c|}{0.5951} & \multicolumn{1}{c|}{27.54} & 0.7254 \\ \hline
K-UNet & \multicolumn{1}{c|}{25.34} & \multicolumn{1}{c|}{0.766} & \multicolumn{1}{c|}{28.76} & \multicolumn{1}{c|}{0.8427} & \multicolumn{1}{c|}{33.96} & 0.9288 & \multicolumn{1}{c|}{27.15} & \multicolumn{1}{c|}{0.6696} & \multicolumn{1}{c|}{28.90} & 0.7710 \\ \hline

Deep ADMM & \multicolumn{1}{c|}{26.17} & \multicolumn{1}{c|}{0.7312} & \multicolumn{1}{c|}{30.41} & \multicolumn{1}{c|}{0.8122} & \multicolumn{1}{c|}{33.94} &\multicolumn{1}{c|}{0.8513}  & \multicolumn{1}{c|}{29.31} & \multicolumn{1}{c|}{0.7515} & \multicolumn{1}{c|}{30.97} &  \multicolumn{1}{c}{0.8162}\\ \hline

SwinMR & \multicolumn{1}{c|}{27.48} & \multicolumn{1}{c|}{0.8893} & \multicolumn{1}{c|}{30.36} & \multicolumn{1}{c|}{0.9341} & \multicolumn{1}{c|}{33.75} &\multicolumn{1}{c|}{0.9659}  & \multicolumn{1}{c|}{28.41} & \multicolumn{1}{c|}{0.7080} & \multicolumn{1}{c|}{30.30} &  \multicolumn{1}{c}{0.7949}\\ \hline

UNet & \multicolumn{1}{c|}{27.63} & \multicolumn{1}{c|}{0.8437} & \multicolumn{1}{c|}{31.27} & \multicolumn{1}{c|}{0.9024} & \multicolumn{1}{c|}{35.02} & 0.9388 & \multicolumn{1}{c|}{28.33} & \multicolumn{1}{c|}{0.7120} & \multicolumn{1}{c|}{31.35} & 0.8283 \\ \hline

D5C5 & \multicolumn{1}{c|}{30.51} & \multicolumn{1}{c|}{0.9241} & \multicolumn{1}{c|}{35.91} & \multicolumn{1}{c|}{0.9743} & \multicolumn{1}{c|}{42.72} & 0.9927 & \multicolumn{1}{c|}{29.78} & \multicolumn{1}{c|}{0.7507} & \multicolumn{1}{c|}{32.00} & \textbf{0.8399} \\ \hline
OUCR & \multicolumn{1}{c|}{31.61} & \multicolumn{1}{c|}{0.9442} & \multicolumn{1}{c|}{36.78} & \multicolumn{1}{c|}{0.9797} & \multicolumn{1}{c|}{42.80} & 0.9933 & \multicolumn{1}{c|}{\textbf{29.95}} & \multicolumn{1}{c|}{0.7522} & \multicolumn{1}{c|}{\textbf{32.04}} & 0.8382 \\ \hline
Ours & \multicolumn{1}{c|}{\textbf{32.58}} & \multicolumn{1}{c|}{\textbf{0.9529}} & \multicolumn{1}{c|}{\textbf{37.47}} & \multicolumn{1}{c|}{\textbf{0.9823}} & \multicolumn{1}{c|}{\textbf{43.68}} & \textbf{0.9946} & \multicolumn{1}{c|}{29.88} & \multicolumn{1}{c|}{\textbf{0.7534}} & \multicolumn{1}{c|}{31.97} & 0.8389 \\ \hline
\end{tabular}%
}
\vspace{-5pt}
\caption{Quantitative comparison on Gaussian 2D sampling settings.}
\label{tab:gaussian}
\vspace{-10pt}
\end{table}

%% file: conclusion-bmvc.tex
\section{Conclusion}

We propose \model, 
a novel Transformer-based architecture for undersampled MRI reconstruction. 
It exploits implicit representation, 
and learns the global dependencies between frequency bins in k-space.
Specifically,
we design a flexible hierarchical decoder to produce high quality reconstruction with moderate computational cost.
Extensive experiments show that the proposed \model~outperforms existing state-of-the-art methods or achieves comparable performance on two public datasets, demonstrating promising results of Transformer in MRI reconstruction and k-space learning. 
Future studies can be conducted to adapt it to multi-coil imaging, k-space super-resolution and other potentially relevant directions.

%% file: Ack.tex
% \vspace{-8pt}
\section*{Acknowledgments}
% \vspace{-10pt}

This work is supported by the Science and Technology Commission of Shanghai Municipality (STCSM) (No. 18DZ2270700, No. 22511106100), 111 plan (No. BP0719010),  and State Key Laboratory of UHD Video and Audio Production and Presentation.

%% file: Appendix.tex
In this supplementary mateirial, we start by introducing implement details on the proposed K-Space Transformer and baseline methods included in our experiments. Then, we demonstrate ablation study on the influence of a deeper K-Space Transformer and how the hybrid learning works in the HR Decoder. Finally, we supplement more qualitative comparison results on settings that are not included in Section 4.2 .

\section{Implement Details}

\textbf{K-Space Transformer}
% Backbone architecture
is implemented with 4 Encoder layers, 
4 Low-Resolution Decoder layers and 6 High-Resolution Decoder layers. 
Each encoder layer is equipped with a four-head self-attention layer and a feed forward network;
each LR decoder layer is equipped with a four-head cross-attention layer, a four-head self-attention layer and a feed forward network; 
and each HR decoder layer is equipped with a four-head cross-attention, a feed forward network and a refinement module.
The backone of refinement module consists of 5 convolution layers with leaky ReLU activations between. The first four layers has 64 filters of 3$\times$3 kernel; while the last layer has 2 filters of 3$\times$3 kernel. Necessary prediction, embedding and fft/ifft layers are inserted before and after the convolutions for transformation between k-space and image domain.
By default, the embedding layers, prediction layers and feed forward networks mentioned in this paper are two-level MLPs with ReLu activations between.
% Dimensionality
Following previous works~\cite{acnn2021,OUCR,ksl2019,D5C5,HybridCascade}, 
we treat the input and output complex MR signals as double channels, {\em i.e.}, real and imaginary. Features vectores and 2D positional encodings mentioned in Section 3 are set with the same dimensionality $d = 256$.

% Loss function
We adopt pixel-wise loss between reconstruction $\mathcal{I}$ and groundtruth $\mathcal{I}_{gt}$ in image domain, and apply deep supervision~\cite{lee2015deeply} on each layer of LR and HR decoders :  $\mathcal{L}=\sum_{i}\|\mathcal{I}_{i} - \mathcal{I}_{gt} \|_2$, where $i$ indicates the layer depth. 
% Training details
We train K-Space Transformer with AdamW optimizer and cosine annealing schedule, setting the initial learning rate as $5\times10^{-4}$. Inspired by ~\cite{PreNorm}, 
we adopt Pre-LN architecture to save warm-up stage. 
As the HR reconstruction depends on the LR decoder output (refer to Section 3.3), 
we find it helpful to stop the gradients from HR decoder outputs in early epochs, {\em i.e.}, only update the LR decoder. Training is conducted on 4 RTX 3090.  \\

\par{\noindent \bf Other Baseline Approaches. } 
We implement a standard U-Net~\cite{UNet} with data consistency layer before the final output for U-Net and K U-Net, which follows the same architecture as ~\cite{ksl2019}. We train the network with the same objective function and learning rate as our methods but adopt an Adam optimizer. We implement D5C5~\cite{D5C5} according to its official configuration for 2D images reconstruction: a cascade network of 5 CNNs, each equipped with 5 layers and a data consistency layer. We train it with the same hyper-parameters as U-Net. For Deep ADMM~\cite{DeepADMM}, SwinMR~\cite{SwinMR} and OUCR~\cite{OUCR}, we adopt their official source codes and default hyper-parameters that are publicly available or provided by the authors.

\section{Influence of Network Depth}

\begin{figure}[!htb]
    \centering
    \includegraphics[width=0.99\textwidth]{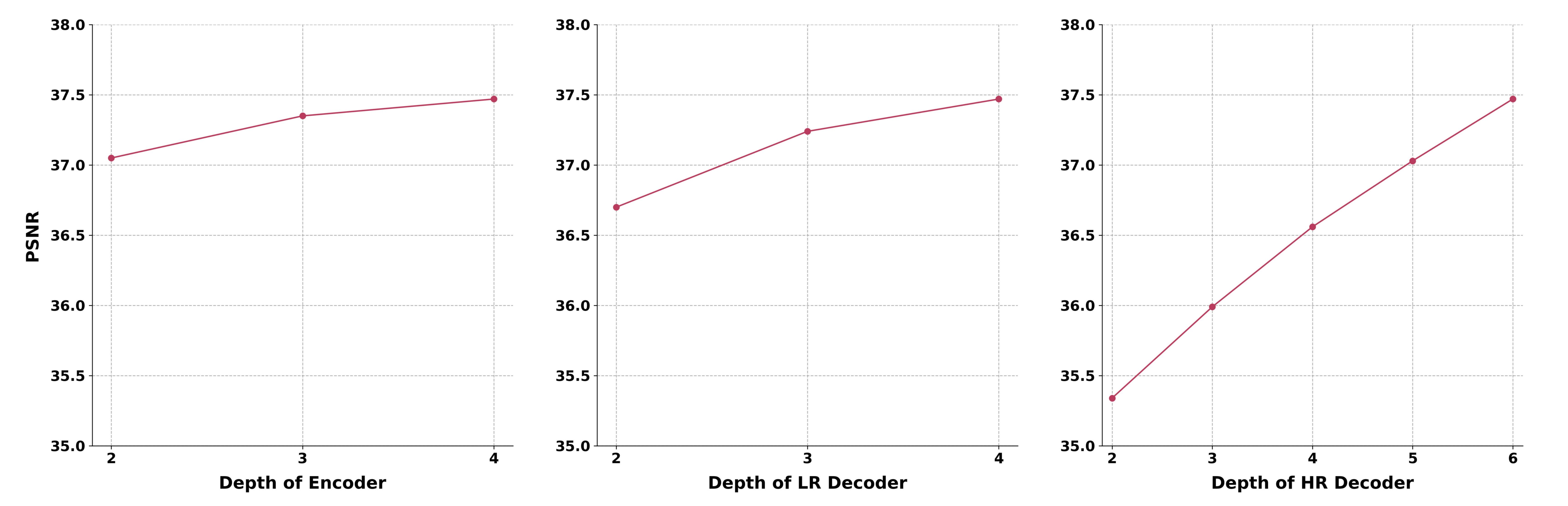}
    \vspace{-0.1cm}
    \caption{The performance influence of increasing module depth.}
    %\vspace{-0.3cm}
    \label{fig:ModelDepth}
\end{figure}

We investigate the effect of layer depth in K-Space Transformer by increasing the depth of Encoder, 
LR Decoder and HR Decoder respectively and evaluate under Gaussian sampling, 
5x acceleration. As demonstrated in Figure~\ref{fig:ModelDepth}, 
increasing depth consistently improve the network performance in all three modules. Though we conjecture our model may benefit from even deeper architecture, we take a default setting as 4 Encoder layers, 
4 Low-Resolution Decoder layers and 6 High-Resolution Decoder layers 
to balance the computational cost.

\begin{figure}[!htb]
    \centering
    \includegraphics[width=0.99\textwidth]{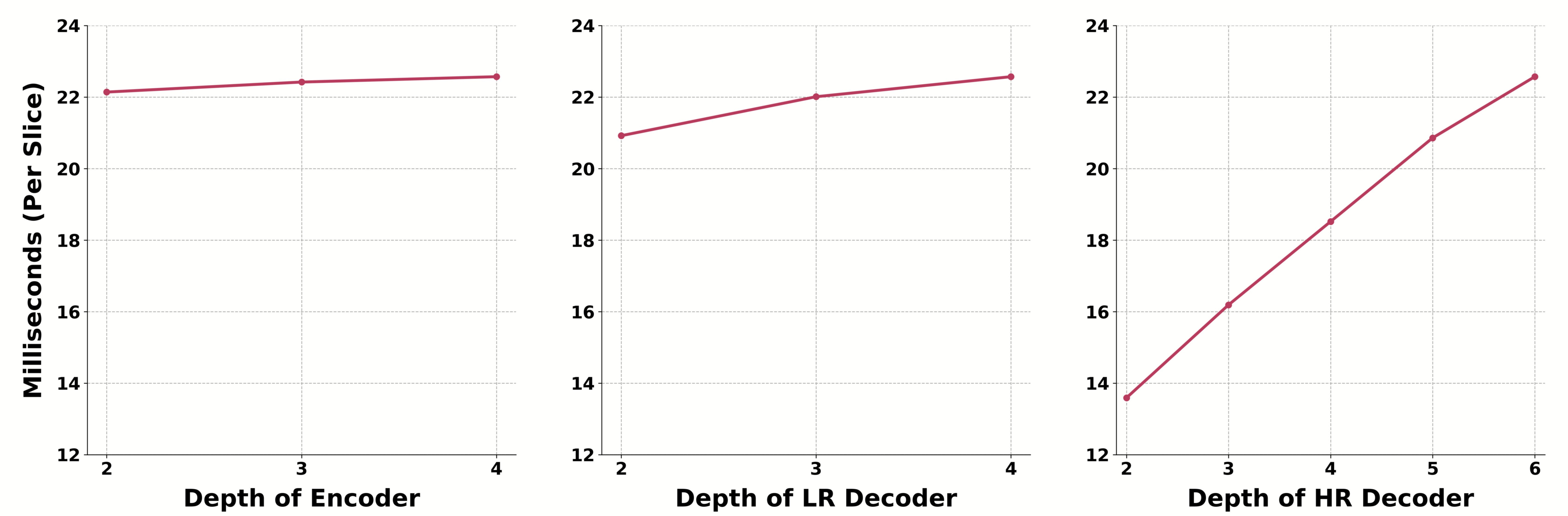}
    \vspace{-0.1cm}
    \caption{The speed influence of increasing module depth.}
    %\vspace{-0.3cm}
    \label{fig:DepthTime}
\end{figure}

\par{\noindent \bf On Reconstruction Speed.} 
Here, we evaluate the reconstruction time, 
as expected, deeper networks tend to incur longer processing time, 
shown in Figure~\ref{fig:DepthTime}. 
For example, depending on the depth of the HR decoder, 
the average inference time for each slice could range from 13.6ms to 22.6ms. 
By comparison, it's significantly faster than Deep ADMM (around 1s) and traditional CS-based methods (up to hundreds of seconds) and comparable to SwinMR (around 18ms), and only slightly slower than the existing CNN-based methods (a few milliseconds). However, we believe the processing speed of our model is sufficient for real-time imaging, thus its remarkable performance improvement on high acceleration settings should be more valuable to improve the patients' scanning experience and reduce medical cost in clinical application.

\section{Effectiveness of Hybrid Learning}

\begin{figure}[!htb]
    \centering
    \includegraphics[width=0.99\textwidth]{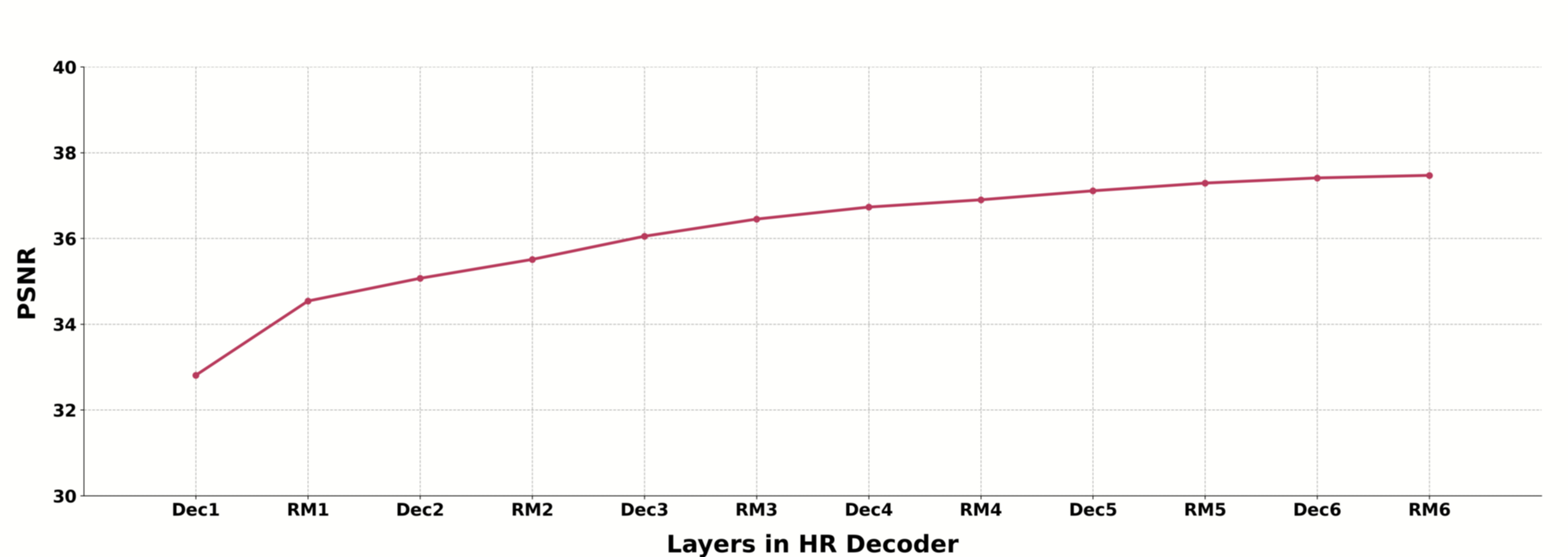}
    \vspace{-0.1cm}
    \caption{Evaluation on deep supervised reconstruction from each k-space decoding(denoted as Dec) and image refinement module(denoted as RM) in the HR Decoder.}
    %\vspace{-0.3cm}
    \label{fig:DecLayers}
\end{figure}

We have shown that image domain refinement does provide complementary information to k-space decoding in Section 4.3 . Here, we further investigate how the alternating process works in the hybrid learning. After evaluating the deep supervised reconstructed results of each HR decoder layer under Gaussian sampling, 5x acceleration (shown in
Figure~\ref{fig:DecLayers}), we observe that PSNR increases consistently throughout the alternation between k-space decoding and image domain refinement, and this finding is consistent across all the settings. 
%As we do not observe oscillating reconstruction quality in deeper layer as \cite{HybridCascade} did, we conjecture this might be owing to that we take a deep supervision strategy and add skip connection on dense feature of k-space decoding in each image refinement module, {\em i.e.}, the \texttt{Query}. 

%\zzh{Not certain about this conjecture: they(alternating k-space CNN and image CNN) observe the loss may oscillate in deeper reconstruction blocks but we didn't reproduce this baseline on our dataset}

\section{Supplementary Qualitative Comparison with Baselines}

\begin{figure}[!htbp]
    \centering
    \includegraphics[width=0.99\textwidth]{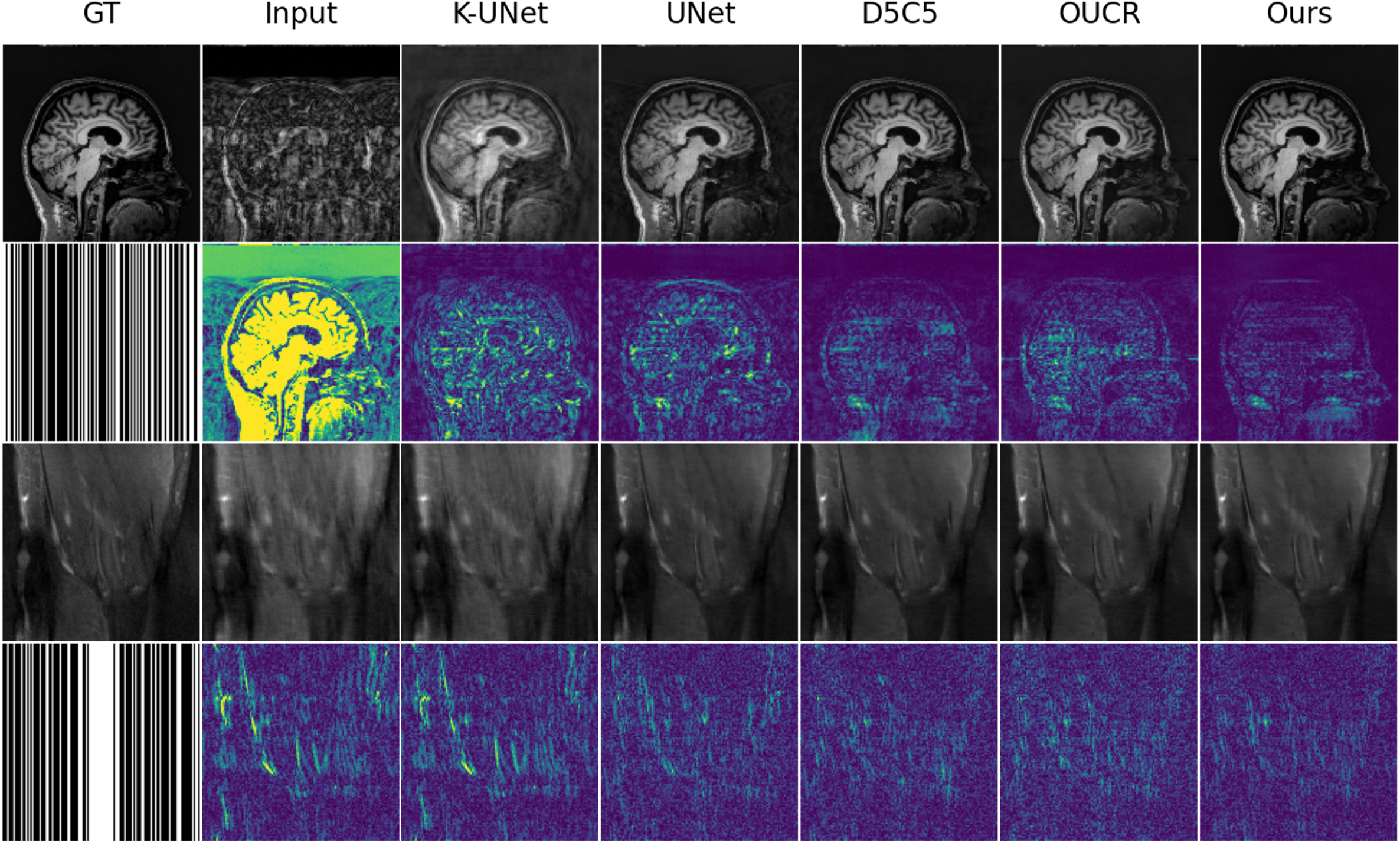}
    \caption{Supplementary qualitative comparison on uniform sampling settings.}
    \label{fig:visual-sup-1}
    % \vspace{-0.3cm}
\end{figure}

\begin{figure}[!htbp]
    \centering
    \includegraphics[width=0.99\textwidth]{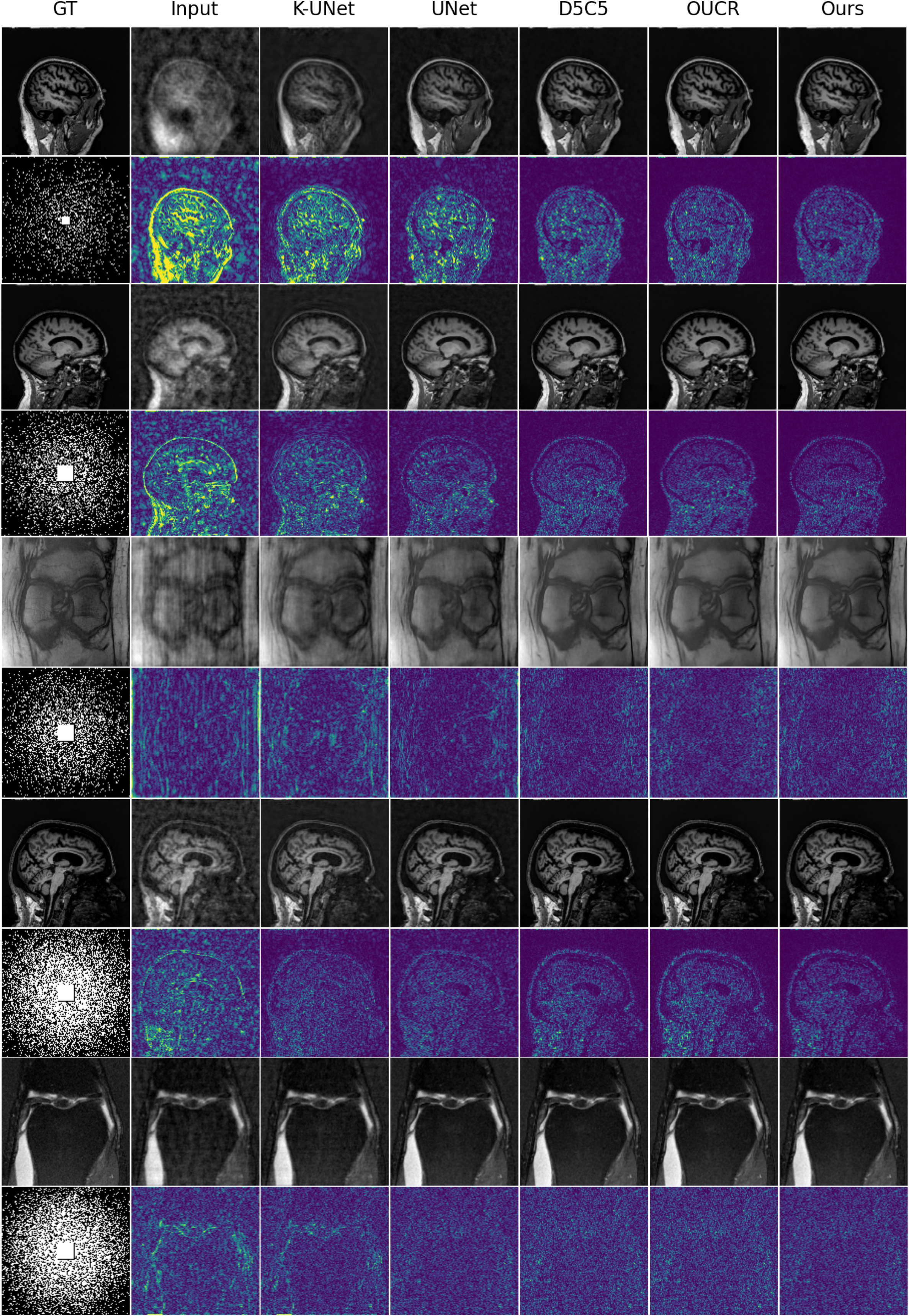}
    \caption{Supplementary qualitative comparison on Gaussian sampling settings.}
    \label{fig:visual-sup-2}
    % \vspace{-0.3cm}
\end{figure}

We show more qualitative comparisons that are not included in Section 4.2 . As can be seen from Figure~\ref{fig:visual-sup-1} and Figure~\ref{fig:visual-sup-2}, on settings like 2.5x uniform sampling or 10x Gaussian sampling, our method achieve remarkably higher reconstruction quality over other approaches, demonstrating the robustness to more significant aliasing artifacts. While on settings with less structure loss and aliasing artifacts, the top methods perform quite close.

